%% file: paper.tex
\documentstyle[prb,aps,amsfonts,amstex,amssymb,floats,twocolumn,exscale,epsfig]{revtex}

\begin{document}

\hyphenation{ma-cro-ionic}

\title{A Monte-Carlo approach to Poisson-Boltzmann like free energy functionals}
\author{Markus Deserno}
\address{Max-Planck-Institut f{\"u}r Polymerforschung, Ackermannweg 10, 55128 Mainz, Germany}
\date{October 19, 1999}
\maketitle


\hspace{2cm}\begin{abstract}
\begin{centering}
\parbox{16cm}{
\vspace{-1cm}
\hfill
\parbox{14.2cm}{
  A simple technique is proposed for numerically determining equilibrium ion
  distribution functions belonging to free energies of the Poisson-Boltzmann
  type.  The central idea is to perform a conventional Monte-Carlo simulation
  using the free energy as the ``Hamiltonian'' entering the Metropolis
  criterion and the spatially discretized density as degrees of freedom.  This
  approach is complementary to the possibility of numerically solving the
  differential equations corresponding to the variational problem, but it is
  much easier to implement and to generalize. Its utility is demonstrated in
  two examples: valence mixtures and hard core interactions of ions
  surrounding a charged rod.
}
}
\end{centering}
\end{abstract}



\narrowtext


\newcommand{\rd}{{\operatorname{d}}}
\newcommand{\re}{{\operatorname{e}}}
\newcommand{\rB}{{\operatorname{B}}}
\newcommand{\rM}{{\operatorname{M}}}
\newcommand{\new}{{\operatorname{new}}}
\newcommand{\old}{{\operatorname{old}}}
\newcommand{\eff}{{\operatorname{eff}}}
\newcommand{\FV}{{\operatorname{FV}}}


\section{Introduction}

\noindent
Understanding the interaction between charged macroions in solution requires
knowledge of how the ions from the surrounding electrolyte screen the
macroionic charge. A common starting point for the description of such systems
is a mean field theory, which expresses the free energy as a functional of the
ionic density. Accounting only for the electrostatic energy and the entropy of
this density field yields (after functional minimization) the nonlinear
Poisson-Boltzmann (PB) equation, which can be solved exactly in certain
special cases \cite{GouyChapman,cylcellmodel,DeHoMa,TracyWidom}. Since this
approach neglects statistical correlations as well as the finite size of the
small ions, various methods have been suggested which approximately account
for those phenomena by adding correction terms to the free energy functional
\cite{correlcorr,DeHoBa,BoAnOr97,LuZoBl99}. However, the analytical treatment
of those extended Poisson-Boltzmann theories is even harder than that of the
original one; consequently, the desired ionic density profile is computed by
numerically solving the complicated differential- or even integro-differential
equations.

In this paper a method is proposed for determining the equilibrium ion
distribution, which completely avoids differential equations.  The key idea is
to perform a straightforward Monte-Carlo (MC) simulation of the free energy
functional. Although this is a numerical approach as well, it has the big
advantage that its computational effort is mostly unaffected by the
mathematical complexity of the functional under investigation.  Only recently
similar techniques have successfully been applied to the Onsager free energy
functional from liquid-crystal theory \cite{WilliJack}.


\section{Description of the Monte-Carlo method}

\noindent
This section briefly outlines the physical and mathematical background
and proceeds with describing the Monte-Carlo scheme. For the sake of
concreteness the derivations are presented within the framework of the
cylindrical cell model, which is a commonly used approach for reducing
the complicated many body problem of a solution of stiff
polyelectrolytes (like, e.g., DNA) to an effective one polyelectrolyte
theory \cite{cylcellmodel}.  Upon {\em (i)\/} assuming that the
surface of zero electric field surrounding one such molecule is on
average a cylinder and {\em (ii)\/} neglecting edge effects by taking
the length of the molecule to be infinite, one arrives at the geometry
depicted in Fig.~{\ref{fig:cell}.
%
%
%
\begin{figure}[t]
\vspace{4.23cm}
\begin{center} \includegraphics{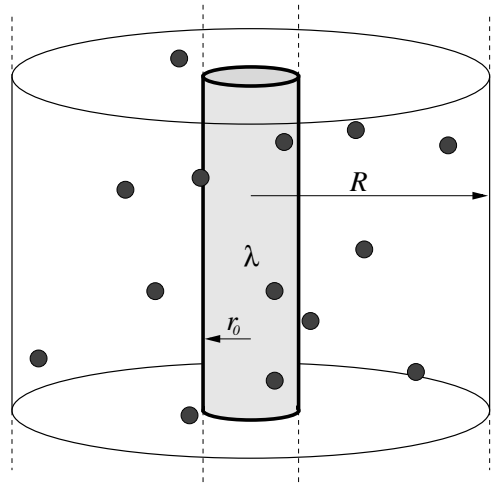} \end{center}
\caption{\sloppy Geometry of the cell model. An infinitely long cylinder
      of radius $r_0$ and line charge density $\lambda>0$ is coaxially
      enclosed in a cylindrical cell of radius $R$. Global charge
      neutrality of the system is ensured by adding an appropriate
      amount of oppositely charged counterions. The outer radius $R$
      is used to match the polyelectrolyte density of a solution
      containing many such macromolecules.}\label{fig:cell}
\end{figure}
%
A convenient measure for electrostatic interaction strength is the {\em
  Bjerrum length} $\ell_\rB = \beta e_0^2/4\pi\varepsilon$ ($k_\rB T =
\beta^{-1}$ being the thermal energy and $\varepsilon$ the dielectric
constant), which permits the Coulomb energy for two unit charges $e_0$
a distance $d$ apart to be written as $k_\rB T \, \ell_\rB / d$.
Within a mean field density functional approach the counterions
surrounding the rod are replaced by a cylindrically symmetric ion
density $n(r)$ ($r$ is the radial coordinate) which gives rise to a
likewise symmetric electrostatic potential $\psi(r)$.  The simplest
ansatz for the free energy restricts to the electrostatic part of the
energy and the entropy:
\begin{equation}
  \hspace*{-2ex}
  F[n(r)] = L\int_{r_0}^R \rd r \, 2\pi r 
  \left\{\frac{\varepsilon}{2}\big(\nabla\psi\big)^2 
    + \frac{n(r)}{\beta}\ln\frac{n(r)}{\bar n}\right\}.
  \label{eq:free_energy_functional}
\end{equation}
Here, $\bar{n}$ is the average counterion density and $L$ is the
length of a cylinder subsegment, with respect to which all extensive
quantities are to be measured in the following.  Functional
minimization of $F$ under the constraint of global charge neutrality
leads to the nonlinear Poisson-Boltzmann equation
\begin{equation}
  \Delta \psi(r) =
  \psi''(r)+\frac{1}{r}\,\psi'(r) =
  n(R) \re^{\beta e_0 \psi(r)},
  \label{eq:PB}
\end{equation}
which has to be solved subject to the boundary conditions
\begin{equation}
  \hspace*{-3ex}
  \psi'(r_0) = -\frac{\lambda}{2\pi\varepsilon r_0}
  \;,\quad
  \psi'(R) = 0
  \quad\text{and}\quad
  \psi(R) = 0.
  \label{eq:PB_boundary}
\end{equation}
The analytical solution of (\ref{eq:PB},\ref{eq:PB_boundary}) and in
particular a discussion of the phenomenon of {\em Manning condensation}
\cite{Manning}, which occurs for a charge parameter $\xi=\lambda\ell_\rB/e_0$
larger than $1$ can e.g.\ be found in Refs.~\cite{cylcellmodel,DeHoMa}. The
present paper suggests a method for (numerically) finding the distribution
$n(r)$ which minimizes functionals like (\ref{eq:free_energy_functional}) {\em
  without\/} using their corresponding differential equations.

Imagine the cell being subdivided by the $M+1$ (not necessarily equidistant)
splitting points $s_0, s_1, \ldots, s_M$ (with $s_0=r_0$ and $s_M=R$) into $M$
concentric cylindrical shells of volume $V_i = \pi(s_i^2-s_{i-1}^2)L$.  If
furthermore the $N_i$ ions within shell $i$ are replaced by a density $n_i =
N_i / V_i$, the set $\{N_i\}$ already completely specifies the (macro-)state
of the system.  The total electrostatic energy $E$ of this configuration can
be computed exactly by piecewise integration of the Poisson
equation in cylindrical geometry, yielding
\begin{eqnarray}
  E(\{N_i\}) & \; = \; & \frac{1}{4\pi\varepsilon}\sum_{i=1}^M\bigg\{
  \big(Q_{i-1}-e_0n_i \pi s_{i-1}^2 L\big)^2
  \ln\frac{s_i}{s_{i-1}} \nonumber \\
  {} & \; {} \; & + \;\; e_0N_i\Big(Q_{i-1} + e_0n_i \,
  \pi\frac{s_i^2-3s_{i-1}^2}{4}L\Big)\bigg\},\label{eq:en_bin_exact}
\end{eqnarray}
where $Q_{i-1}$ is the total charge up to but not including shell $i$.  The
idea now is to perform a Markov process by randomly moving particles between
the shells and accepting the moves with the usual Metropolis criterion
\cite{Metropolis}. In the canonical ensemble the probability of a state
$\{N_i\}$ is proportional to the product of the Boltzmann factor $P_E =
\exp\{-\beta E\}$ and the probability $P_S$ of distributing the
(indistinguishable) ions between the shells in a way compatible with the given
state, i.e.,
\begin{equation}
  P_S\big(\{N_i\}\big) \; = \;
  \frac{N!}{V^N} \prod_{i=1}^{M} \frac{V_i^{N_i}}{N_i!}
  \label{eq:P_S}
\end{equation}
where $N=\sum_i N_i$ and $V=\sum_i V_i$. Generate a {\em new} state $\{N_i'\}$
from the {\em old} one by moving one ion from shell $k$ to shell $l$. Since
\begin{equation}
  \frac{P_S\big(\{N_i'\}\big)}{P_S\big(\{N_i\}\big)} \; = \;
  \frac{N_k}{V_k}\frac{V_l}{N_l+1} \; = \;
  \frac{n_k^\old}{n_l^\new},\label{eq:PS_ratio}
\end{equation}
detailed balance between the total probabilities finally yields the following
acceptance probability of such a move:
\begin{equation}
  \hspace*{-3ex}
  \text{Prob}\big(\,k\stackrel{1}{\longrightarrow} l\,\big) = 
  \min\left\{1,\;\frac{n_k^\old}{n_l^\new}\re^{-\beta\big(E^\new\!-E^\old\big)}
    \right\}
  \label{eq:MCPB_Metropolis}
\end{equation}
There is an alternative way of looking at this: Since the entropy of a state
is given by $S = k_\rB\ln P_S$, the expression entering in
(\ref{eq:MCPB_Metropolis}) can actually be written as $\exp\{-\beta F\}$,
i.e., the ``combinatoric'' multiplicity of the states is automatically taken
into account if the Metropolis criterion is tested with the {\em free\/}
energy.  Furthermore, if all $N_i$ are large, the factorials in
(\ref{eq:P_S}) can be approximated by Stirling's formula:
\begin{equation}
S\big(\{N_i\}\big) \; \simeq \; -k_\rB\,V\, 
\sum_{i=0}^M n_i \ln\frac{n_i}{\bar{n}}.\label{PBMC_entropy}
\end{equation}
This is the shell-discretized equivalent of the entropy contribution entering
into (\ref{eq:free_energy_functional}). By performing this Markov process one
can thus sample the ion distribution function corresponding to the free energy
functional (\ref{eq:free_energy_functional}) (after an initial equilibration
time). 

\vspace{2ex}

\noindent
Some final remarks:
\begin{enumerate}
\item The proposed technique is applicable to all geometries in which the
  electrostatic energy can be computed efficiently once the charge density in
  the volume elements is known. In particular, this applies to the spherical
  and cylindrical cell model as well as the charged plane, because their high
  symmetry permits equations like (\ref{eq:en_bin_exact}).
\item Still referring to {\em particles} within a density functional theory
  might seem artificial: one could equally well transfer small bits of the
  {\em density} between shells (under conservation of the integral) and employ
  $\exp\{-\beta F\}$ in the Metropolis criterion with the entropy contribution
  calculated from (\ref{PBMC_entropy}). However, Stirling's approximation is
  only good for large $N_i$, therefore, the global extensive prefactor in the
  exponential function must be large enough -- for otherwise one even fails to
  reproduce a constant density in the limit of zero charge. Since in the {\em
    spherical} geometry the size of this prefactor is not at ones disposition
  (the total number of counterions per colloid is proportional to its finite
  and possibly small charge), equation (\ref{eq:P_S}) is to be preferred to
  equation (\ref{PBMC_entropy}).
\item Upon measuring the electrostatic interactions via the Bjerrum length,
  the explicit temperature dependence drops out of the Metropolis criterion.
\item Further contributions $F'$ to the free energy functional (e.g., excluded
  volume or correlation corrections) can be included by adding corresponding
  terms $-\beta\,\Delta F'$ in the exponential function of
  (\ref{eq:MCPB_Metropolis}).
\item The approach bears some resemblance with the method of {\em finite
    elements}, for which the transformation from a differential equation to a
  functional minimization problem is a central idea \cite{HeBo96}.
\end{enumerate}


\begin{figure}[t]
  \begin{minipage}{\textwidth}
    \begin{minipage}{0.68\textwidth}
      \input{fig2}
    \end{minipage}
    \hfill
    \begin{minipage}{0.31\textwidth}
      \vspace*{-0.5cm}
      \caption{Poisson-Boltzmann charge fractions $P(r)$ (solid lines) from
        equation (\ref{eq:P(r)}) for a valence mixture system with $R/r_0=100$,
        $\lambda=\frac{4}{3}\,e_0/r_0$ and $\ell_\rB/r_0=1$.  The lowest curve
        corresponds to 100\% monovalent counterions, while for the uppermost curve
        all ions are trivalent. From bottom to top the monovalent ions are
        gradually replaced by trivalent ones in steps of 10\% of the charge.  The
        dashed line indicates the locus of inflection points in $P(r)$ as a
        function of $\ln(r)$ for the PB theory with only {\em one\/} species of
        ions (in this case for varying $\xi$), whereas the heavy dots mark the
        {\em actual\/} inflection points in the distribution
        functions.}\label{fig:val_mix}
    \end{minipage}
  \end{minipage}
\end{figure}
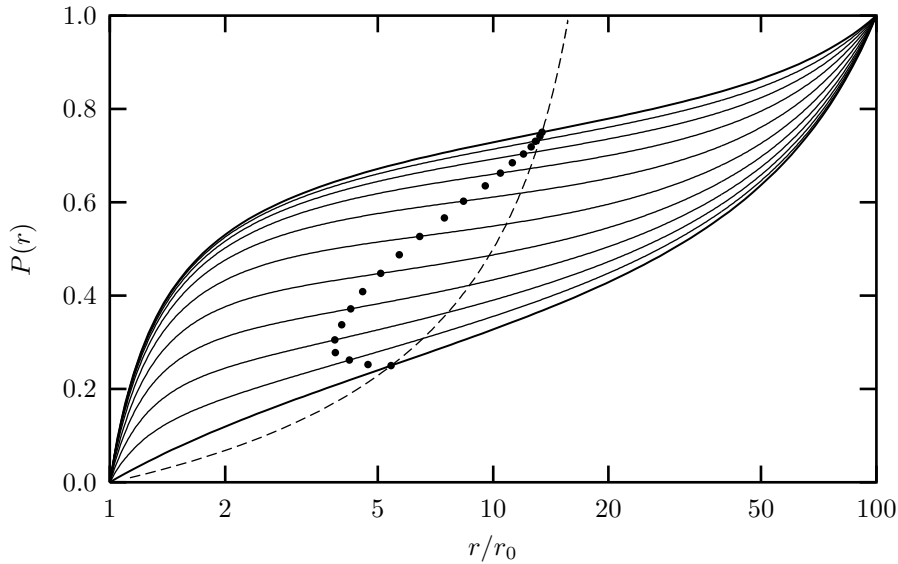

\vspace*{-0.1cm}

\section{Exemplary applications}

\noindent
In order to demonstrate the utility of the described technique, this section
presents two possible applications: Within the cell model described above the
ionic distribution functions are computed for {\em (i)\/} a system with
ions having different valences and {\em (ii)\/} a system with
additional hard core interactions.


\subsection{Valence mixtures}

\noindent
The PB equation in the cylindrical cell model cannot be solved analytically
for a system containing a {\em mixture\/} of ions having different valences.
For the MC approach, however, this is no problem, since it extends in an
obvious way to such a situation: Each species of valence $v_j$ is represented
by its own density $n_j(r)$ and a corresponding histogram.  MC moves
only act within the same histogram and the electrostatic energy is computed
from the total charge density $\rho(r)=e_0\sum_j v_j n_j(r)$.
Figure~\ref{fig:val_mix} shows successive stages of an ion exchange process:
The fraction of charge within a distance $r$ from the rod axis,
\begin{equation}
  P(r) \; = \; \frac{1}{\lambda}\int_{r_0}^r \! \rd\bar{r} \, 
  2\pi \bar{r} \, \rho(\bar{r}),
  \label{eq:P(r)}
\end{equation}
is plotted for a system with $R/r_0=100$, $\lambda=\frac{4}{3}\,e_0/r_0$ and
$\ell_\rB/r_0=1$, in which the neutralizing monovalent counterions are gradually
(i.e., in steps of 10\% of the charge) replaced by trivalent ones -- until
100\% of the charge is carried by the latter. The cell was subdivided into
$500$ shells with boundaries equidistant in $\ln r$, $10^6$ MC moves were
performed to equilibrate the system and on average $5 \times 10^7$ further
moves were used for sampling the distribution. The number of particles which
were in minority ranged between $200$ and $700$.

%
\begin{figure}[t]
  \vspace{7.65cm}
\end{figure}

As expected, the condensed fraction gets larger with increasing fraction of
trivalent ions, i.e., the distribution functions are shifted upwards.
Interestingly though, there is also a more subtle effect connected with the
screening of the rod: It has recently been shown that within PB theory the
point of inflection in $P(r)$ as a function of $\ln(r)$ localizes the
condensation radius and the corresponding condensed fraction \cite{DeHoMa}. If
only one species of counterions is present, the locus of inflection points in
$P(r)$ (parameterized by $\xi$) is solely determined by the cell geometry. In
Fig.~\ref{fig:val_mix} it can be seen that for the mixtures the {\em actual\/}
inflection points are shifted towards {\em smaller\/} radii, i.e., the rod
charge is screened more efficiently. This effect must be attributed to {\em
  global rearrangements\/} of the counterions; the distribution of one species
clearly depends on the distribution of the other, different valences are
correlated -- even on the PB level.  A more detailed analysis in fact reveals
that highvalent ions will gather in the vicinity of the rod at the expense of
lowvalent ones.  This provides a way of further decreasing the free energy,
therefore, mixtures can screen more efficiently (i.e., within shorter
distances) than solutions of only one valence. Notice finally that the
observed shift implies that there is no ``effective'' valence $v_\eff$ such
that a fictitious solution of counterions with valence $v_\eff$ gives the same
distribution function as the actual valence mixture.


\vspace*{0.05cm}

\subsection{Excluded volume}

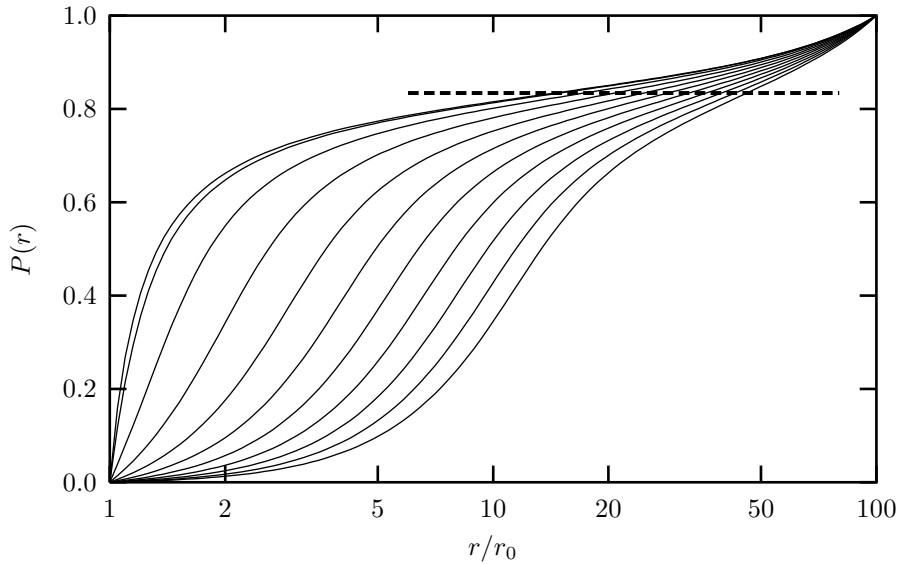
\begin{figure}[t]
  \begin{minipage}{\textwidth}
    \begin{minipage}{0.68\textwidth}
      \input{fig3}
    \end{minipage}
    \hfill
    \begin{minipage}{0.28\textwidth}
      \vspace*{-0.5cm}
      \caption{Counterion fraction $P(r)$ for
        a cylindrical cell model with $R/r_0=100$, $\lambda=2\,e_0/r_0$,
        $\ell_\rB/r_0=1$ and $v=3$. The diameter $d$ of the ions (implemented via
        (\ref{eq:free_volume_f}) with $n_{\max}=3/2\pi d^3$) has been varied from
        $0$ to $10\,r_0$ in steps of $r_0$, which shifts the initial rise in
        $P(r)$ to larger values of $r$. The straight dashed line is a fit to the
        locus of inflection points relevant for the condensation
        argument.}\label{fig:ex_vol}
    \end{minipage}
  \end{minipage}
\end{figure}

\noindent
The free energy functional (\ref{eq:free_energy_functional}) of the plain
Poisson-Boltzmann equation assumes no interaction beyond
electrostatics. In particular, it neglects all effects resulting from an ionic
hard core, which range from a suppression of high densities up to a packing of
particles. A simple way for incorporating at least the density limitation is
the {\em free-volume approximation}, characterized by the following
contribution to the free energy density:
\begin{equation}
  \beta f_\FV(n) \; = \; - n \, \ln \big( 1-n/n_{\max} \big),
  \label{eq:free_volume_f}
\end{equation}
where $n_{\max}$ is the maximally allowed density.  (A related approach has
recently been used to incorporate steric repulsion in the PB equation, see
Ref.~\cite{BoAnOr97}.) After including (\ref{eq:free_volume_f}) into
(\ref{eq:free_energy_functional}), an MC simulation for a cell model has been
performed with $R/r_0=100$, $\lambda=2\,e_0/r_0$, $\ell_\rB/r_0=1$ and $v=3$.
As the limiting density $n_{\max}=3/2\pi d^3$ has been chosen (with
$d/r_0\in\{0,1,\dots,10\}$), which reproduces the correct second virial
coefficient for hard spheres of diameter $d$.  $500$ shells and $2000$
particles have been used for $d=r_0$, while for systems with larger ions the
number of shells has been reduced and the number of particles increased ($50$
shells and $50000$ particles for $d=10\,r_0$) in order to avoid discretization
effects occurring in the small innermost shells. After $200$--$500$ MC steps
per particle for equilibration, a further $5000$--$10\,000$ per particle have
been used to sample the distributions. The effects on $P(r)$ are shown in
Fig.~\ref{fig:ex_vol}.

The distribution function for $d=r_0$ deviates only slightly from the one with
$d=0$. The reason for this is that the contact density in the absence of a
hard core, $n(r_0)=0.447\,r_0^{-3}$, is of the same order as the maximum
density $n_{\max,d=r_0}=0.477\,r_0^{-3}$. However, larger values of $d$
bring about the expected changes: the contact density is reduced in the
vicinity of the rod since the ions are pushed outwards. Therefore the sharp
initial rise of $P(r)$ shifts to larger values of $r$ and softens out.
Surprisingly though, the {\em amount\/} of condensation (for $d=0$ one expects
$1-1/\xi v \approx 83\%$ \cite{Manning}) is mostly unaffected by the
pronounced changes in the ion distribution at small $r$.  This can be seen
from the straight line fit to the locus of relevant inflection points:
Although the condensed layer considerably increases in size (from $14.3\,r_0$
up to $52.1\,r_0$), the thus quantified condensed fraction decreases by less
than one percent.


%
\begin{figure}[t]
  \vspace{7.6cm}
\end{figure}

\section{Conclusions}

\noindent
A method for finding the equilibrium ion distribution belonging to a free
energy functional of the Poisson-Boltzmann type has been proposed. It avoids
any reference to the differential equation corresponding to the variational
problem, which makes it very easy to adapt to new free energy functionals. Its
utility has been demonstrated in two examples, but its range of applicability
is clearly much larger: It can also be used in the presence of salt, with
additional free energy contributions (like, e.g., correlation corrections
\cite{DeHoBa}) and in different geometries. In fact, the basic idea could be
useful in the general context of functional minimization.


\begin{acknowledgments}
\noindent
The author would like to thank B.\ D\"unweg and C.\ Holm for many inspiring
discussions.
\end{acknowledgments}


\end{document}

%% file: fig2.tex
\setlength{\unitlength}{0.1bp}
\begin{picture}(3239,2160)(0,0)
\special{psfile=fig2 llx=0 lly=0 urx=648 ury=504 rwi=6480}
\put(1794,50){\makebox(0,0){$r/r_0$}}
\put(50,1180){%
\special{ps: gsave currentpoint currentpoint translate
270 rotate neg exch neg exch translate}%
\makebox(0,0)[b]{\shortstack{$P(r)$}}%
\special{ps: currentpoint grestore moveto}%
}
\put(3239,200){\makebox(0,0){100}}
\put(2804,200){\makebox(0,0){50}}
\put(2229,200){\makebox(0,0){20}}
\put(1795,200){\makebox(0,0){10}}
\put(1360,200){\makebox(0,0){5}}
\put(785,200){\makebox(0,0){2}}
\put(350,200){\makebox(0,0){1}}
\put(300,2060){\makebox(0,0)[r]{1.0}}
\put(300,1708){\makebox(0,0)[r]{0.8}}
\put(300,1356){\makebox(0,0)[r]{0.6}}
\put(300,1004){\makebox(0,0)[r]{0.4}}
\put(300,652){\makebox(0,0)[r]{0.2}}
\put(300,300){\makebox(0,0)[r]{0.0}}
\end{picture}

%% file: fig3.tex
\setlength{\unitlength}{0.1bp}
\begin{picture}(3239,2160)(0,0)
\special{psfile=fig3 llx=0 lly=0 urx=648 ury=504 rwi=6480}
\put(1794,50){\makebox(0,0){$r/r_0$}}
\put(50,1180){%
\special{ps: gsave currentpoint currentpoint translate
270 rotate neg exch neg exch translate}%
\makebox(0,0)[b]{\shortstack{$P(r)$}}%
\special{ps: currentpoint grestore moveto}%
}
\put(3239,200){\makebox(0,0){100}}
\put(2804,200){\makebox(0,0){50}}
\put(2229,200){\makebox(0,0){20}}
\put(1795,200){\makebox(0,0){10}}
\put(1360,200){\makebox(0,0){5}}
\put(785,200){\makebox(0,0){2}}
\put(350,200){\makebox(0,0){1}}
\put(300,2060){\makebox(0,0)[r]{1.0}}
\put(300,1708){\makebox(0,0)[r]{0.8}}
\put(300,1356){\makebox(0,0)[r]{0.6}}
\put(300,1004){\makebox(0,0)[r]{0.4}}
\put(300,652){\makebox(0,0)[r]{0.2}}
\put(300,300){\makebox(0,0)[r]{0.0}}
\end{picture}

%% file: paper.bbl
\begin{thebibliography}{999}
%
\bibitem{GouyChapman} G.\ Gouy, J.\ Phys. (France) 9 (1910) 457; D.\ L.\ 
  Chapman, Philos.\ Mag.\ 25 (1913) 475.
%
\bibitem{cylcellmodel} T.\ Alfrey, P.\ Berg and H.\ J.\ Morawetz, J.\ Polym.\ 
  Sci.\ 7 (1951) 543. R.\ M.\ Fuoss, A.\ Katchalsky and S.\ Lifson, Proc.\ 
  Natl.\ Acad.\ Sci.\ USA, 37 (1951) 579. M.\ Le Bret and B.\ H.\ Zimm,
  Biopolymers 23 (1984) 287.
%
\bibitem{DeHoMa} M.\ Deserno, C.\ Holm and S.\ May, submitted to {\em
    Macromolecules}.
%
%
\newpage
%
%
\bibitem{TracyWidom} C.\ A.\ Tracy and H.\ Widom, Physica A 244 (1997) 402.
%
\bibitem{correlcorr} S.\ Nordholm, Chem.\ Phys.\ Lett.\ 105 (1984) 302. R.\ 
  Penfold, S.\ Nordholm, B.\ J\"onsson and C.\ E.\ Woodward, J.\ Chem.\ Phys.
  92 (1990) 1915. R.\ D.\ Groot, J.\ Chem.\ Phys. 95 (1991) 9191.
%
\bibitem{DeHoBa} M.\ C.\ Barbosa, M.\ Deserno and C.\ Holm, to be published.
%
\bibitem{BoAnOr97} I.\ Borukhov, D. Andelman and H.\ Orland, Phys.\ Rev.\ 
  Lett.\ 79 (1997) 435.
%
\bibitem{LuZoBl99} L.\ Lue, N.\ Zoeller and D.\ Blankschtein, Langmuir 15
  (1999) 3726.
%
\bibitem{WilliJack} D.\ C.\ Williamson and G.\ Jackson, Mol.\ Phys.\ 83 (1994) 
  603.
%
\bibitem{Manning} G.\ S.\ Manning, J.\ Chem.\ Phys.\ 51 (1969) 924; 934; 3249.
%
\bibitem{Metropolis} N.\ Metropolis, A.\ W.\ Rosenbluth, M.\ N.\ Rosenbluth,
  A.\ H.\ Teller and E.\ Teller, J.\ Chem.\ Phys.\ 21 (1953) 1087.
%
\bibitem{HeBo96} D.\ Henwood and J.\ Bonet, {\em Finite Elements. A Gentle
    Introduction}. MacMillan, London, 1996.
%
\end{thebibliography}
